# Airborne Radiometric Surveys and Machine Learning Algorithms for Revealing Soil Texture

**Andrea Maino [1,2,*], Matteo Alberi [1,2], Emiliano Anceschi [3], Enrico Chiarelli [1,2], Luca Cicala [4], Tommaso Colonna [5], Mario De Cesare [4,6,7], Enrico Guastaldi [5], Nicola Lopane [1,5], Fabio Mantovani [1,2], Maurizio Marcialis [8], Nicola Martini [9], Michele Montuschi [1,2], Silvia Piccioli [9], Kassandra Giulia Cristina Raptis [1,2], Antonio Russo [3], Filippo Semenza [1,2] and Virginia Strati [1,2]**

[1] Department of Physics and Earth Sciences, University of Ferrara, 44122 Ferrara, Italy; alberi@fe.infn.it (M.A.); chrnrc@unife.it (E.C.); lpnncl@unife.it (N.L.); mantovani@fe.infn.it (F.M.); montuschi@fe.infn.it (M.M.); rptksn@unife.it (K.G.C.R.); fsemenza@fe.infn.it (F.S.); strati@fe.infn.it (V.S.)

[2] INFN Ferrara Section, 44122 Ferrara, Italy

[3] Gruppo Filippetti Sede Falconara Marittima, 60015 Falconara Marittima, Ancona, Italy; emiliano.anceschi@gruppofilippetti.it (E.A.); antonio.russo@gruppofilippetti.it (A.R.)

[4] CIRA, Italian Aerospace Research Centre, 81043 Capua, Caserta, Italy; l.cicala@cira.it (L.C.); m.decesare@cira.it (M.D.C.)

[5] GeoExplorer Impresa Sociale S.r.l., 52100 Arezzo, Italy; colonna@geoexplorersrl.it (T.C.); guastaldi@geoexplorersrl.it (E.G.)

[6] Department of Mathematics and Physics, University of Campania "Luigi Vanvitelli", 81100 Caserta, Italy

[7] INFN Napoli Section, Complesso Universitario di Monte S. Angelo, 80126 Napoli, Italy

[8] Corso Giuseppe Garibaldi 119, 44022 Comacchio, Ferrara, Italy; maurizio.marcialis@gmail.com

[9] Le Due Valli S.r.l., 44020 Ostellato, Ferrara, Italy; nicola.martini@leduevalli.com (N.M.); silvia.piccioli@leduevalli.com (S.P.)

* Correspondence: maino@fe.infn.it

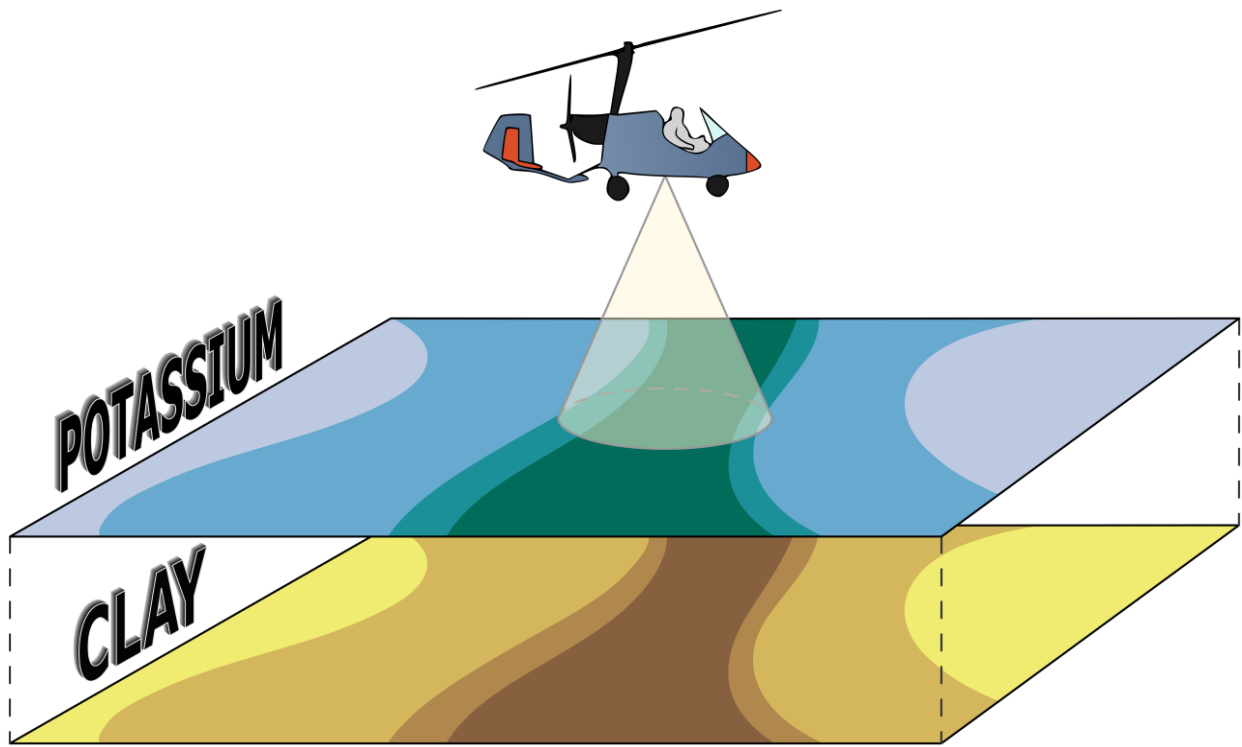

**Abstract:** Soil texture is key information in agriculture for improving soil knowledge and crop performance, so the accurate mapping of this crucial feature is imperative for rationally planning cultivations and for targeting interventions. We studied the relationship between radioelements and soil texture in the Mezzano Lowland (Italy), a 189 km$^2$ agricultural plain investigated through a dedicated airborne gamma-ray spectroscopy survey. The K and Th abundances were used to retrieve the clay and sand content by means of a multi-approach method. Linear (simple and multiple) and non-linear (machine learning algorithms with deep neural networks) predictive models were trained and tested adopting a 1:50,000 scale soil texture map. The comparison of these approaches highlighted that the non-linear model introduces significant improvements in the prediction of soil texture fractions. The predicted maps of the clay and of the sand content were compared with the regional soil maps. Although the macro-structures were equally present, the airborne gamma-ray data permits us shedding light on finer features. Map areas with higher clay content were coincident with paleo-channels crossing the Mezzano Lowland in Etruscan and Roman periods, confirmed by the hydrographic setting of historical maps and by the geo-morphological features of the study area.

**Keywords:** airborne gamma-ray spectroscopy; non-linear machine learning; potassium; clay; thorium; sand; soil texture; paleo-hydrography

## 1. Introduction

Soil is an essential resource that, now more than ever, plays a multifaceted and crucial role in human well-being and society. Its regulating functions on global climate and water availability, as well as food production, provide in fact a fundamental link between soil and life on Earth [1]. The improvement of land quality, the reduction in soil pollution and the restoration of degraded lands are all cross-cutting themes of the 2030 Agenda for sustainable development goals [2]. It follows that the understanding of the complex soil properties and their interaction processes must be a basic principle for the sustainable management of this resource.

In precision farming, the availability of high-resolution maps of the soil's physical and chemical parameters is emerging as a crucial need to reduce the risk of yield failure and improve crop management [3,4]. In this puzzle, soil texture is undeniably the main influencing parameter in several biological, chemical and physical processes. The grain size of sediments affects transport, deposition and erosion processes occurring in soils [5,6], as well as the property of the mineral surface of adsorbing organic matter, nutrients and pollutants [7,8]. Soil texture is a decisive controlling factor of the hydraulic conductivity and hence the water surface infiltration [9], which strongly impacts groundwater availability and flow water paths. Textural features presently observed are valuable records of the evolutional processes involving the filling of abandoned channels resulting from rivers' shifting processes, such as meander cutoff and channel-belt avulsion [10,11].

The quantification of the sand, silt and clay components in the soil is traditionally performed via direct methods (e.g., sieving, sedimentation, laser-diffraction analysis), and scattered measurements with all the drawbacks arising from them, such as the limited size of the investigated volume, time-consuming and destructive operations [12,13]. Indirect measurements of the soil texture through remote sensing surveys can overcome these limitations, providing high-resolution maps of soil properties. In addition, soil reflectance measurements from multi-spectral and radar sensors mounted on board satellites [14-16] and gamma-ray data acquired through proximal surveys have proved to be useful and promising options for predicting soil texture [17-19]. Even if the basic application of gamma-ray spectroscopy in soil science is to investigate the composition of the source bedrock, the gamma signal emitted by soil is likewise related to the abundances of the granulometric classes and to the weathering and pedogenesis processes [20]. Clay and silt particles tend to behave as colloids. Their elevated specific surface area is responsible for the cation's adsorption (e.g., $K^+$, $U^{4+}$, $U^{6+}$ and $Th^{4+}$) that increases with the grade of weathering [21,22]. Focusing on the mineral structure, as a general rule, the natural radioelements are more abundant in clay minerals (typical of the fine fraction) than in quartz (the main constituent of sand) [23]. The application of machine learning algorithms for soil texture estimation has been proven to be an effective analysis technique. The absence of any a priori assumption on the relationship types between soil fractions and radioelement abundances permits overcoming the site specificity arising from the influence of the parent materials in the cation's adsorption [24-26]. The term "radioelement abundance" refers to the mass fraction (weight percentages) of the corresponding radioelement in the gamma-emitting material, i.e., in the soil for this case. K abundances are indicated in $10^{-2}$ g g$^{-1}$, while U and Th abundances are indicated in μg g$^{-1}$.

In this study, the performance of airborne gamma-ray spectroscopy (AGRS) is tested for the discrimination of soil texture properties with a multi-approach method. AGRS data acquired through a dedicated survey in the Mezzano Lowland (Emilia-Romagna, Italy) are analyzed to retrieve the soil textural fractions of the investigated soils via simple linear regression (SLR) and multiple linear regression (MLR) analysis and through the application of non-linear machine learning (NLML) algorithms, a promising technique for the analysis of agro-environmental and geological data remotely acquired [27-29]. The study area (~189 km$^2$) was investigated for the first time with a homogenous coverage and was revealed to be particularly suitable for this research purpose. Firstly, in this flat and rural area that has experienced extensive reclamation actions in the past century, the almost exclusive presence of agricultural soils drastically reduced the influence of anthropic structures on the measured radiometric data. Secondly, the public soil texture map of the Emilia-Romagna Region (RER) at a 1:50,000 scale can be fully exploited for the correlation studies and the training stages of the NLML analysis. Finally, the possibility of comparing AGRS data with hydrographic historical maps is a valuable opportunity to infer important indications about the Po River delta setting evolution in the Mezzano Lowland.

## 2. Materials and Methods

The study area, known as Mezzano Lowland, is an ~189 km$^2$ flat area located in the Emilia-Romagna region (Italy) in the eastern area of the Po plain (Figure 1). The Mezzano Lowland was a brackish marsh until the 1960s, when land reclamation processes converted it into cropland. This area is one of the most sparsely populated in Italy and is characterized by its rural appearance and the scarce presence of buildings and infrastructure. Indeed, at the time of the reclamation, the use of cars was already quite diffuse, and farmers chose to live in countryside villages. The absence of infrastructure greatly reduces the anthropic disturbance in the airborne gamma signal and makes the area particularly suitable for the purpose of this study. Nowadays, the whole area is below sea level (−3.5 to −0.5 m a.s.l.) and is almost entirely constituted by agricultural fields grown with cereals, soy and tomato crops.

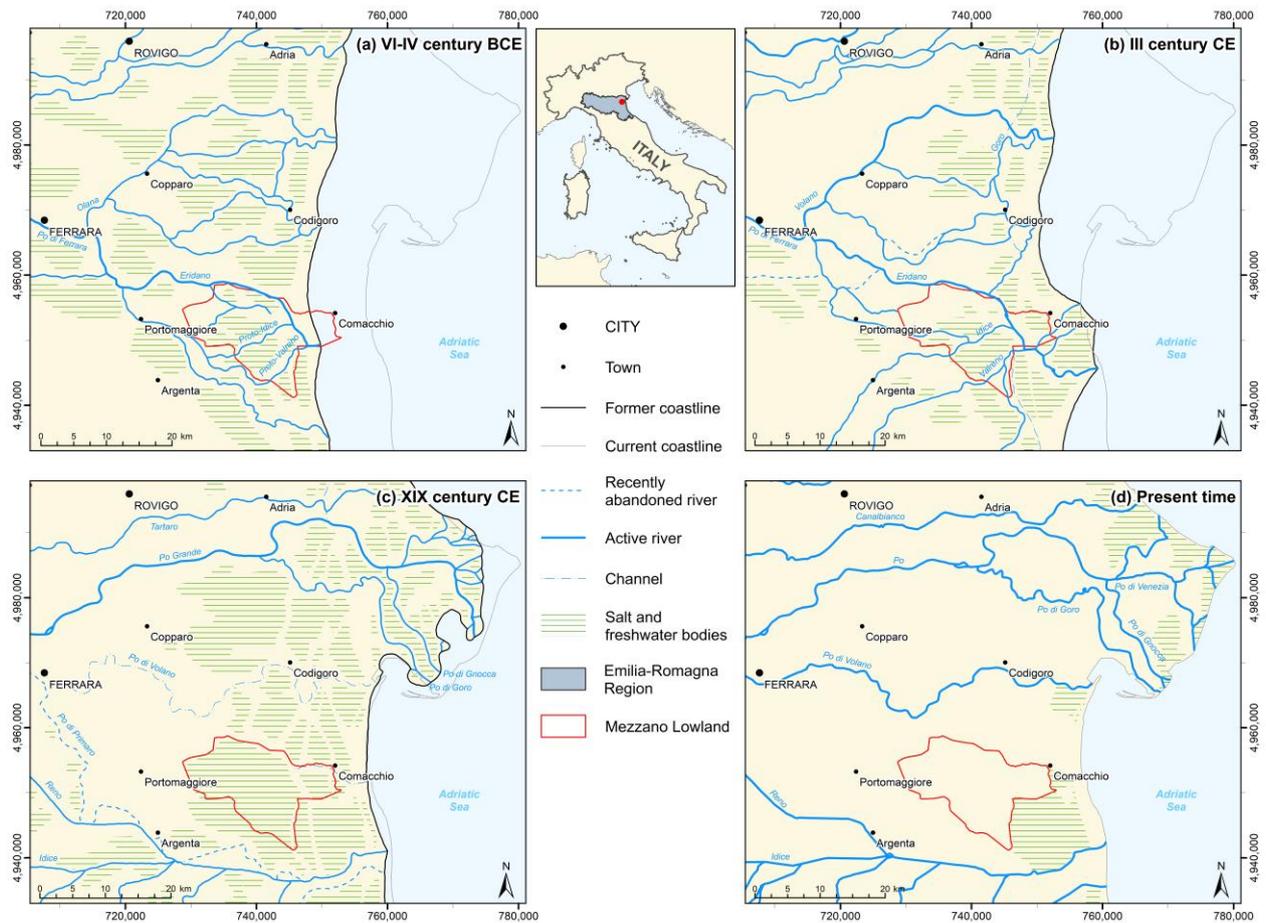

**Figure 1.** The Po plain area near the Mezzano Lowland in the VI-IV century BCE (**a**) III century CE (**b**), XIX century CE (**c**) and its current setting (**d**). For the tracing of the area's evolution, historical maps from local archives [30] were adopted and simplified. For each period, the main hydrographic features and coastline are shown; current placement of cities, towns and coastline, together with the limits of the Mezzano Lowland study area are reported in all panels. Cartographic reference systems: WGS 84 and UTM Zone 32N.

*2.1. Hydrographic Evolution of the Study Area*

The retracing of the hydrographic evolution of the Mezzano Lowland from the Pleistocene (2.58–0.012 Ma) to the Holocene (0.012 Ma—present) is a key point for understanding the current textural features of its soils and their spatial distribution.

The Po plain formed during the Pleistocene when the alluvial sediments, eroded from the Alps and the Apennines Mountain ranges and then transported by the Po River and its tributaries, filled the marine gulf present at the time in the area. The easternmost portion, corresponding to the Po delta, has always been characterized by multiple sedimentation, erosion and subsidence processes strongly influenced by climate variations [30].

The current area of the Mezzano Lowland was crossed, from the X century BCE until the VII century CE, by the hydrographic system "Po di Ferrara—Eridano" (part of the eastern Po river's terminal sector), with the Eridano, Idice and Valreno channels playing a prominent role in the evolution of this area (Figure 1).

Historical records show that during the VI-IV century BCE the Po di Ferrara—Eridano system included the Proto-Idice and Proto-Valreno channels, both originating from a subparallel distributary branching off nearby the current location of the town of Portomaggiore (Figure 1a). Since information about these subsidiary channels are uncertain in the pre-roman period, we named them Proto-Idice and Proto-Valreno to indicate an earlier stage of their evolution process. These channels crossed the area of the Mezzano Lowland from SW to NE, reconverging downstream in the Eridano's riverbed. At this time, the coastline was located approximately 10 km west with respect to the present one. During the III century CE, the Eridano's delta formation caused the eastward migration of the coastline near Comacchio, while artificial fluvial diversions changed the provenance of Idice and Valreno waters, without affecting their path within the Mezzano Lowland (Figure 1b). In the following centuries, the intense erosion processes of the delta and the spread of salt marshes in the whole area led, in the VII century CE, to the complete abandonment of the Po di Ferrara—Eridano system with the consequent sedimentary filling of their riverbeds (Figure 1c). During the XII century CE, the southern branches of the Po delta further lost their importance due to the northward migration of the Po river, which defined a new riverbed (later known as Po Grande) flowing north of the city of Ferrara, with the consequent accretion of another delta and the migration of the surrounding coastline (Figure 1c). The Mezzano Lowland, characterized by diffuse and pervasive salt marshes and by the absence of a hydrographic system (Figure 1c), remained substantially unchanged until the successive land reclamation processes of the XX century (Figure 1d). The reclamation was carried out in the 1960s through the implementation of a pit system, which collects and converges waters into a canal network. It is worth noticing that after 1740, a portion of the Eridano's riverbed was artificially reoccupied by the canal known today as Canale Navigabile, which is part of the current drainage and irrigation system.

*2.2. Soil Texture Data*

The Geological, Seismic and Soil Service of RER provides soil texture maps, relative to the upper 30 cm of soil, at 1:50,000 scale at 500 m × 500 m resolution for the entire regional territory. These maps were obtained by interpolating punctual textural data via the Scorpan Kriging method, with geopedological data serving as ancillary information [31]. In this study, we extracted, for the 723 square meshes belonging to the Mezzano Lowland, the weight percent content of clay, sand, and silt together with the associated texture class defined according to the United States Department of Agriculture (USDA) classification. The reliability class associated with the estimation of sand, silt and clay content is defined as "low" for 84%, 66% and 85% of the meshes, respectively; for the remaining, "medium" or "high" classes are assigned.

*2.3. Radiometric Data*

AGRS is a well-established technique that has been applied in geoscience studies for decades with diverse purposes, ranging from mining exploration to geological mapping [20,32-36].

The AGRS survey was performed using the Radgyro [37,38] (Figure 2a), an aircraft equipped with GPS antennas and altimetric sensors for the logging of the geographic latitude and longitude and the flight altitude [39]. A modular NaI(Tl) scintillation detector, composed with four 4L-crystals, is placed in the middle of the Radgyro hull. The survey was divided into three flights for a total of 4 h and 45 min and 482 km (Figure 2b), with a mean velocity of $102 \pm 13$ km h$^{-1}$ and at a mean height of $104 \pm 21$ m. The acquired 1469 gamma spectra were then analyzed offline to obtain information on Th, K and U ground abundances with a time resolution of 10 s, compared approximately to a measurement every 280 m of flight. The spectra analysis was performed through the application of the full spectrum analysis (FSA) with non-negative least squares [40] which considers the acquired spectrum as a result of a composition of the fundamental spectra of each radionuclide and of the background radiation given by cosmic contribution and intrinsic background of the instrumentation [37,38]. In this case, the fundamental spectra are reconstructed, at the corresponding flight altitude, through a Monte Carlo code based on GEANT 4 [41] for the simulation of the gamma emissions of a radionuclide homogeneously distributed over an infinite source. Following the method described in [38], the atmospheric $^{222}$Rn daughters are estimated to produce an additional effect in the acquired gamma spectra, leading to a potential overestimation of the ground U abundance approximately equal to 1.1 µg g$^{-1}$, 0.9 µg g$^{-1}$ and 0.7 µg g$^{-1}$ in Flight 1, 2 and 3 (Figure 2b), respectively. This non-constant contribution can be explained by the poor ventilation typical of the study area which is below the sea level and with a high relative air humidity. Considering this not negligible bias and the evident lack of secular equilibrium, we decided not to include U content in this study. The minimum detection abundances are $0.05 \times 10^{-2}$ g g$^{-1}$ for K and 0.08 µg g$^{-1}$ for Th [37]. K and Th abundances maps were obtained applying an ordinary kriging interpolation to obtain a 723 square mesh grid with a 500 m × 500 m resolution and coincident with the RER soil texture maps.

For the investigation of areas with dimensions comparable to those of the Mezzano Lowland (~200 km$^2$), AGRS surveys are more convenient than in situ measurements for two reasons: (i) the collection and measurement of 1469 soil samples would have required a huge temporal and economical effort; (ii) the soil samples are in most cases representative of a very small volume (~10 cm$^3$), while AGRS detects gamma signals emitted by an area of ~0.2 km$^2$ integrating local variations, which could introduce biases in the soil texture prediction. The observation scale of AGRS is a good compromise towards an expedient solution to optimize both time and economic resources.

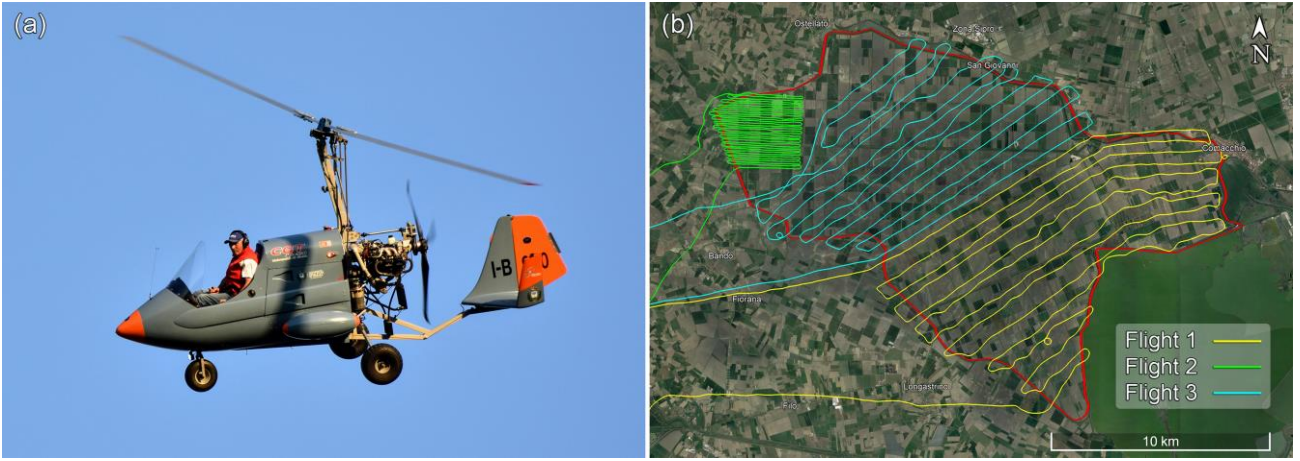

**Figure 2.** (**a**) The Radgyro, the aircraft used for the AGRS surveys and (**b**) paths of Flights 1, 2 and 3, ranging 184 km, 126 km and 172 km in the Mezzano Lowland, respectively. The distance of the flight lines is ~600 m for Flights 1 and 3; the smaller line spacing of Flight 2 (~100 m) arises from the spatial resolution requirements of the photogrammetric survey simultaneously performed.

*2.4. Regression Analysis*

The linear relations between K and Th abundances and sand, silt and clay soil fractions are estimated, firstly, with an SLR analysis and, secondly, with a MLR model. A NLML algorithm was developed for studying more complex relationships. The 723 input data of the entire dataset are split, through a random (non-stratified) sampling, into a training and a testing dataset with a ratio of 80:20 (corresponding to 578:145). The linear and non-linear relations are studied for the training dataset, while the testing dataset is used for validating and comparing the three prediction models. The absence of bias in data selection, proved by comparable statistical parameters of the two datasets, demonstrated that the adopted sampling method permitted having representative sets of the input data.

The SLR model assumes that textural fraction values depend on either K or Th abundance through the equation:

$$Y\ [\%] = m \cdot a(X) + q, \quad (1)$$

where Y stands for clay, silt or sand weight fraction, $a(X)$ is the abundance of the radioelement X (in $10^{-2}$ g g$^{-1}$ for K or in µg g$^{-1}$ for Th), and m and q are the model's parameters to be determined. The uncertainties on the estimations of m and q parameters (Table 1) are given by their standard deviations $\delta m = \sqrt{\sum_{i=1}^{N}(Y_i - \widehat{Y}_i)^2 / \left[(N-2) \cdot \sum_{i=1}^{N}(a(X)_i - \overline{a(X)})^2\right]}$ and $\delta q = \sqrt{\left(\sum_{i=1}^{N}(Y_i - \widehat{Y}_i)^2 / N - 2\right) \cdot \left(1/N + \overline{a(X)}^2 / \sum_{i=1}^{N}(a(X)_i - \overline{a(X)})^2\right)}$, where N is the total number of data, $Y_i$ is the i-th soil texture fraction value provided by RER, $\widehat{Y}_i$ is the i-th predicted soil texture fraction value, $a(X)_i$ is the i-th measured abundance of radioelement X and $\overline{a(X)}$ is the mean value of the $a(X)_i$. The linear correlation is investigated using the Pearson correlation coefficient r defined as:

$$r = \frac{\sigma_{Ya(X)}}{\sigma_Y \sigma_{a(X)}}, \quad (2)$$

where $\sigma_{Ya(X)}$ is the covariance between the soil fraction Y and the radioelement abundance $a(X)$ and $\sigma_Y$, $\sigma_{a(X)}$ are the respective standard deviations.

In MLR the relations between clay, silt, or sand content and both K and Th abundances are assumed as:

$$Y\ [\%] = a \cdot a(K)\,[10^{-2}\ g\ g^{-1}] + b \cdot a(Th)\,[\mu g\ g^{-1}] + c, \quad (3)$$

where a, b and c are model parameters that must be determined. The uncertainties da, db and dc reported in Table 3 are calculated by taking the square root of the diagonal elements of $\sigma^2 (\mathbf{X}^T\mathbf{X})^{-1}$, where $\sigma^2 = 1/(N - p - 1) \cdot \sum_{i=1}^{N}(Y_i - \widehat{Y}_i)^2$ is the variance of the errors, N is the number of data, p is the number of independent variables, $Y_i$ is the i-th observed value of the dependent variable and $\widehat{Y}_i$ is the i-th prediction of the model, while $\mathbf{X}$ is the matrix of the independent variables at each of the N data points. The linear dependence of the textural fraction on K and Th abundances is quantified via the multiple correlation coefficient R as:

$$R = \left| \sqrt{1 - \frac{\sum_{i=1}^{N} e_i^2}{\sum_{i=1}^{N}(Y_i - \overline{Y})}} \right|, \quad (4)$$

Where N is the number of observations, $Y_i$ is the observed value, $\overline{Y}$ is the average value of the $Y_i$ and $e_i$ is the i-th error given by $e_i = Y_i - \hat{Y}_i$ where $\hat{Y}_i$ is the predicted value. The statistical significance of the a, b, and c parameters is assessed by means of a *t*-test with a *p*-value based on the null hypothesis of non-significance in the multi-linear relation.

Going beyond the linear fit of a single radioisotope (K or Th) and clay or sand content, a NLML algorithm was developed by composing two identical supervised deep neural networks tuned for predicting clay and sand soil contents, using both K and Th abundances as input features (dimensionality = 2). The "Adam" optimizer [42] was set by the algorithm to minimize the mean squared error between the outputs and RER soil texture data (target labels) over 100 epochs with a fixed learning rate of 0.001. The model is designed using 6 layers (Figure 3): the first (Input layer) is a normalization layer using Keras' standard normalization function [43], the following four are fully-connected hidden layers with 16 nodes activated by the rectified linear unit function (ReLU) [44], and the last one (Output layer) is a linear single-output layer. The hyperparameters adopted (i.e., learning rate, number of epochs, density and number of hidden layers, optimizer chosen, optimization parameter and activation function) have been tuned by trial and error to avoid under- and over-fitting the models. The small intrinsic variability of the clay and sand content predictions on the testing dataset is reduced by taking the mean values for the outputs over 100 runs of the algorithm. Finally with the purpose of estimating USDA soil textural classes, the silt content was calculated as silt [%] = 100 − clay [%] − sand [%].

Since the three regressions were calculated with 578 data (training dataset), the remaining 145 values (testing dataset) are used for investigating the models' predictive performance studying the coefficient of determination $R^2$ defined as:

$$R^2 = 1 - \frac{\sum_{i=1}^{N}(Y_i - \hat{Y}_i)^2}{\sum_{i=1}^{N}(Y_i - \overline{Y})^2}, \tag{5}$$

where N, $Y_i$, $\hat{Y}_i$, $\overline{Y}$ are as in Equation (4).

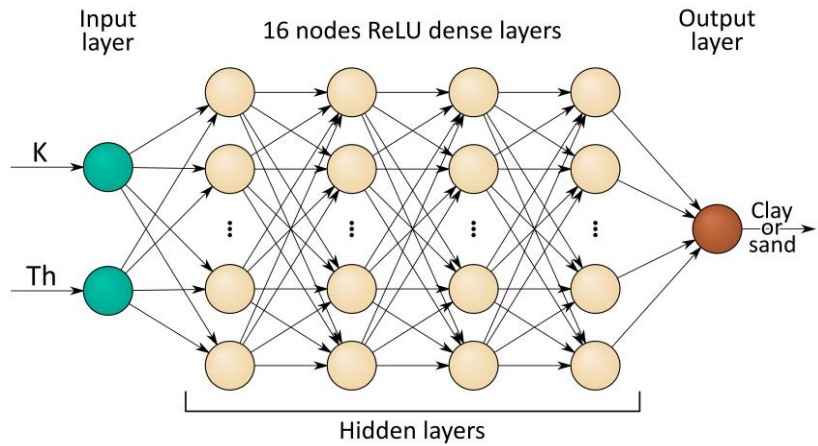

**Figure 3.** Deep neural network model diagram for predicting the output (clay or sand weight fraction) starting from the input features (K and Th abundances). The hidden layers include four dense (fully connected) layers with 16 nodes activated by the ReLU function.

## 3. Results

Adopting the data reported in the RER soil texture maps, the mean clay, silt and sand contents in the Mezzano Lowland are 26 ± 9%, 37 ± 7% and 37 ± 14%, respectively; the predominant texture class is loam, followed by clay loam and sandy loam. The radiometric data (training dataset) showed, for K and Th abundances, a low negative skewness (−0.6) with mean values of a(K) = 0.93 ± 0.14 × $10^{-2}$ g g$^{-1}$ and a(Th) = 6.9 ± 0.9 μg g$^{-1}$ (Figure 4). The Pearson correlation coefficient r = 0.82 indicates a strong linear correlation among the radioelement abundances (Table 1), confirmed also by the low standard deviation of their ratio (a(Th)/a(K) = 7.52 ± 0.75 × $10^{-4}$). Since the silt content has no evident correlation with K and Th abundances (r = 0.47 and r = 0.48), the linear regression models and their implications are not considered in the following discussions.

**Table 1.** Results of the SLR analysis between textural fractions (clay, silt and sand) and radioelements abundances (a(K) and a(Th)) performed on the training dataset (N = 578). The parameters (m and q) of the SLR model are reported together with their standard deviations (δm and δq) and the Pearson correlation coefficient (r); the highest r values for relations between soil fraction and radioelement abundance are shown in italic.

|  |  | a(K) [$10^{-2}$ g g$^{-1}$] | a(Th) [µg g$^{-1}$] | Clay [%] | Silt [%] |
|---|---|---|---|---|---|
| a(Th) [µg g$^{-1}$] | m ± δm | 5.0 ± 0.1 [$10^{-4}$ g g$^{-1}$] |  |  |  |
|  | q ± δq | 2.3 ± 0.1 [µg g$^{-1}$] |  |  |  |
|  | r | 0.82 |  |  |  |
| Clay [%] | m ± δm | 39.9 ± 2.0 [g g$^{-1}$] | 5.4 ± 0.4 [$10^4$ g g$^{-1}$] |  |  |
|  | q ± δq | −11.5 ± 1.9 [%] | −11.6 ± 2.5 [%] |  |  |
|  | r | *0.64* | 0.53 |  |  |
| Silt [%] | m ± δm | 22.2 ± 1.7 [g g$^{-1}$] | 3.7 ± 0.3 [$10^4$ g g$^{-1}$] | 0.5 ± 0.0 [g g$^{-1}$] |  |
|  | q ± δq | 16.5 ± 1.6 [%] | 11.7 ± 2.0 [%] | 24.8 ± 0.7 [%] |  |
|  | r | 0.47 | 0.48 | 0.64 |  |
| Sand [%] | m ± δm | −62.1 ± 3.2 [g g$^{-1}$] | −9.1 ± 0.6 [$10^4$ g g$^{-1}$] | −1.5 ± 0.0 [g g$^{-1}$] | −1.8 ± 0.0 [g g$^{-1}$] |
|  | q ± δq | 95.0 ± 3.0 [%] | 99.9 ± 3.9 [%] | 75.2 ± 0.7 [%] | 106.0 ± 1.6 [%] |
|  | r | *−0.62* | −0.56 | −0.93 | −0.88 |

The clay and sand fractions are plotted as a function of K (Figure 4a,b) and Th (Figure 4c,d) abundances and are then fitted with an SLR model. The 68% prediction level bands demonstrate that the obtained predictive models work better with loam and clay loam soils, while the worst estimations are obtained for values of sand > 60% or of clay > 35%.

The SLR analyses show that the clay content has a moderate positive correlation with K (r = 0.64) and Th (r = 0.53) abundances (Table 1). The sand content has the opposite trend with a moderate negative correlation with K (r = -0.62) and Th (r = −0.56) abundances (Table 1). These results must be attributed to the low capacity of the main constituent of sand (quartz) to adsorb cations and, therefore, radioelements [23,45]. The findings of the correlation analysis substantially agree with similar studies performed with smaller datasets and on different soil types (Table 2). The only exceptions regard Spadoni and Voltaggio [19], which highlight an absence of correlation between a(K) and clay fraction (r = 0.05), and Petersen [18], which reported a negative correlation (r = −0.42) between the two. Authors of the latter study explain this trend as a consequence of "the mobility of [K] ion due to the small radius compared to that of U and Th" [18] (p. 655), despite stating in the same paper that the clay content "shows high positive correlation with cation exchange capacity" [18] (p. 654, Figure 3).

The intercept (q) of the SLR models between sand content and radioelement abundances is compatible at the 2σ level with q = 100%, indicating that a soil with a high sand content (>95%) has K and Th abundances close to zero.

**Table 2.** Comparison of the Pearson correlation coefficient (r) found in this study with the results of similar surveys for which the mean clay and sand contents of the investigated soils are reported together with the number of analyzed data.

| | | | | r | | | |
|---|---|---|---|---|---|---|---|
| | N° of Data | Clay [%] | Sand [%] | Clay vs. K | Sand vs. K | Clay vs. Th | Sand vs. Th |
| This study | 578 | 26 | 37 | 0.64 | −0.62 | 0.53 | −0.56 |
| Van Der Klooster, et al. [46] * | 53 | 18 | / | 0.56 | / | 0.63 | / |
| Mahmood, et al. [47] ** | 36 | 19 | 63 | / | / | 0.62 | −0.51 |
| Spadoni and Voltaggio [19] | 21 | 27 | 21 | 0.05 | −0.48 | 0.07 | −0.07 |
| Elbaalawy, et al. [48] | 16 | 24 | 52 | 0.61 | −0.71 | / | / |
| Petersen [18] | 13 | 25 | 36 | −0.42 | <−0.10 | 0.53 | −0.78 |

* Values reported for the NMD—salt marsh basin—Mn..A soil series characterized by calcareous soils and similar to the study area (Table 4 of the reference). ** Values reported using the FSA method for the combined fields.

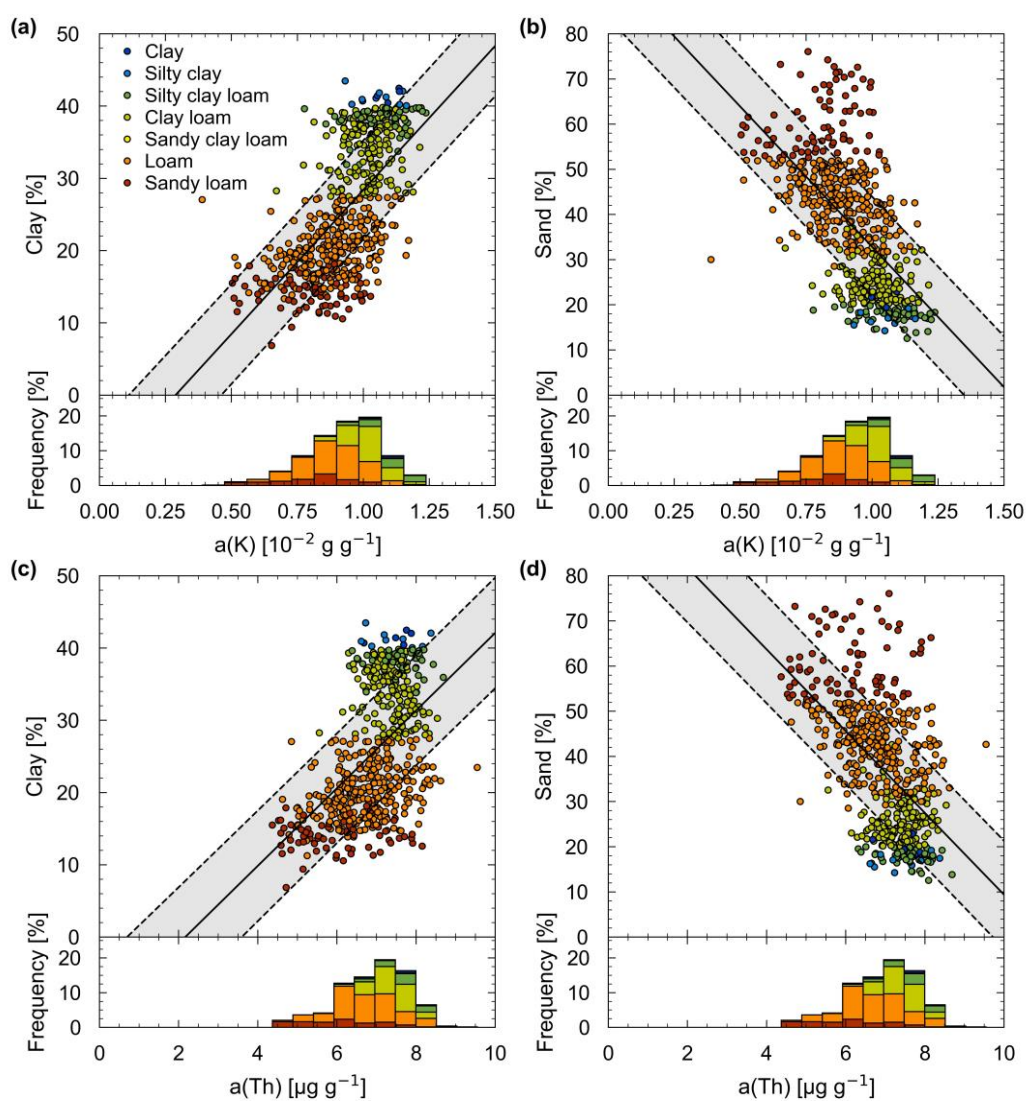

**Figure 4.** Correlation plots between (**a**) clay content and a(K), (**b**) sand content and a(K), (**c**) clay content and a(Th), (**d**) sand content and a(Th) of the training dataset (N = 578) paired with frequency distribution histograms of the corresponding radioelement abundance. Solid lines represent the SLR models with their 68% prediction level bands (gray).

From the MLR analysis results, the sand and clay fractions appear to be moderately correlated with the radioelement abundances (R > 0.60), while the silt content has a lower correlation grade (R = 0.50) (Table 3). The parameter b of the predictive model of the clay content is compatible with zero, highlighting a negligible contribution of the Th content. Testing for the statistical significance of the b parameter in the MLR, we found $p$-values of $9 \times 10^{-1}$, $2 \times 10^{-5}$ and $2 \times 10^{-2}$, respectively, for clay, silt and sand relations, pointing towards a non-significance at the 0.01 level for the contribution of Th to the determination of clay and sand contents.

**Table 3.** Results of the MLR analyses between textural fractions (clay, silt and sand) and radioelement abundances (a(K) and a(Th)) performed on the training dataset (N = 578). The parameters (a, b and c) are reported together with their associated uncertainties (δa, δb and δc) and the multiple correlation coefficient, R.

| Relation | a ± δa [g g$^{-1}$] | b ± δb [$10^4$ g g$^{-1}$] | c ± δc [%] | R |
|---|---|---|---|---|
| Clay [%] = a × a(K) [$10^{-2}$ g g$^{-1}$] + b × a(Th) [μg g$^{-1}$] + c | 39.4 ± 3.5 | 0.1 ± 0.6 | −11.7 ± 2.3 | 0.64 |
| Silt [%] = a × a(K) [$10^{-2}$ g g$^{-1}$] + b × a(Th) [μg g$^{-1}$] + c | 11.6 ± 3.0 | 2.1 ± 0.5 | 11.7 ± 2.0 | 0.50 |
| Sand [%] = a × a(K) [$10^{-2}$ g g$^{-1}$] + b × a(Th) [μg g$^{-1}$] + c | −51.0 ± 5.6 | −2.2 ± 0.9 | 100.0 ± 3.7 | 0.63 |

The soil textural fractions predicted by the MLR model for the testing dataset clearly follow a straight line when plotted in the USDA soil textural triangle (Figure 5a), resulting in more accuracy for the loam and clay loam classes and less reliability for sandy loam and silty clay loam. The soil textural fractions predicted by the NLML algorithm plotted in the same diagram (Figure 5b), not being bound to any linear constraint, seem to better adapt to the scattered RER data; only the silty clay, clay and sandy clay loam classes, which are associated to only 5% of the testing dataset of RER, do not have any NLML predictions.

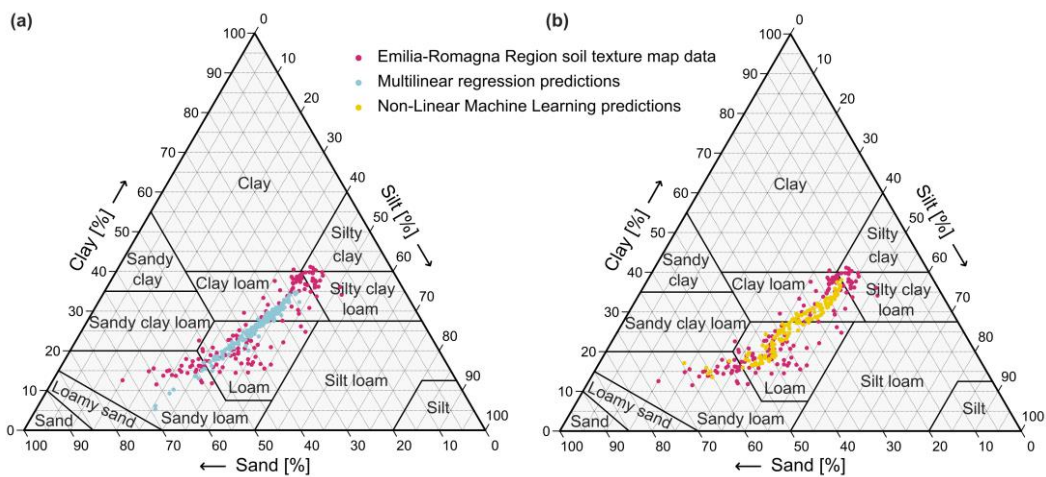

**Figure 5.** USDA soil textural triangle with data extracted from the RER map (145 data, testing dataset) together with (**a**) the corresponding values on the basis of soil textural fractions predicted by adopting MLR models of sand and clay content and (**b**) the corresponding values obtained on the basis of soil textural fractions predicted with the NLML algorithm.

The performances of the three approaches (SLR, MLR and NLML) are compared by means of scatter plots (predicted vs. observed values) and calculation of $R^2$ (Figure 6). The angular coefficient m of the regression line "Predicted value = m × Observed value" is close to 1 (0.92 ≤ m ≤ 0.94) for each predictive model. The highest values of $R^2$ ($R^2$ = 0.52 for clay and $R^2$ = 0.49 for sand) are obtained with NLML (Figure 6e,f) highlighting that the non-linear models perform better compared to the linear ones. At the same time, the results confirm that the MLR does not introduce any significant improvements in terms of m and $R^2$ (Figure 6c,d) with respect to SLR (Figure 6a,b).

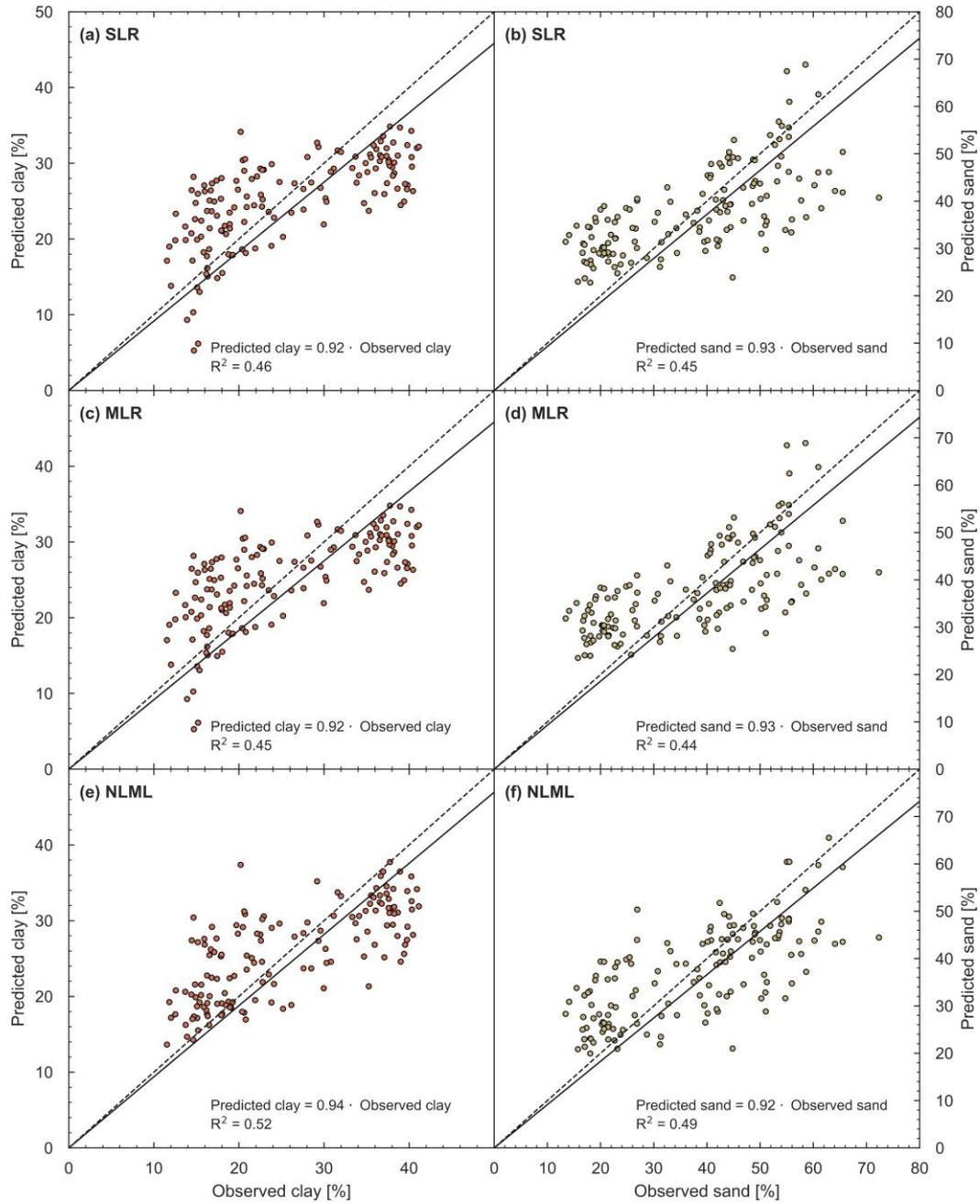

**Figure 6.** Scatter plot of the observed clay values in the RER soil texture map and predicted values by (**a**) the SLR model starting from K abundances, (**c**) the MLR model and (**e**) the NLML algorithm. Scatter plot of the observed sand values in the RER soil texture map and predicted values by (**b**) the SLR model starting from K abundances, (**d**) the MLR model and (**f**) the NLML algorithm. For each plot, the equation of the regression line (solid line) is reported together with the coefficient of determination ($R^2$); the dashed line corresponds to the bisector (Predicted values = Observed values).

The performances of the linear (SLR and MLR) and non-linear (NLML) models were investigated by also studying the spatial distribution of the predicted values, building the predicted maps of the clay and of the sand content, and comparing them with the RER maps. In the following discussion, the MLR maps are not taken into account given their almost identical results to SLR; all the considerations made for the SLR model are likewise valid in the case of MLR.

Adopting the predictive models derived from the SLR analysis:

$$\text{clay } [\%] = 39.9 \cdot a(K) \ [10^{-2} \ g \ g^{-1}] - 11.5, \tag{6}$$

$$\text{sand } [\%] = -62.1 \cdot a(K) \ [10^{-2} \ g \ g^{-1}] + 95.0, \tag{7}$$

the clay and sand content values for each of the 723 meshes are obtained starting from the K abundances quantified via the AGRS measurements (Figure 7c,d). Note that the predictive models based on K abundances are preferred over models based on Th abundances due to the higher values of r. The same maps are obtained by assigning to the 723 meshes the clay (Figure 7e) and sand (Figure 7f) content values predicted by the NLML algorithm.

The maps resulting from the SLR and NLML approaches exhibit the same macrostructures characterized by high values of clay in the north and high values of sand in the central-SE area of the RER soil texture maps (Figure 7a,b), but show finer details. The maps of the differences between clay contents predicted by the SLR (Figure 8a) and NLML (Figure 8b) regression models and those reported by the RER permit the investigation of the spatial distribution of the datapoints furthest from the bisector in Figure 6a,e. The high clay content in the northern belt, associated with the negative anomalies in Figure 8, corresponds to the Eridano's riverbed reported in Figure 9 where the hydrographic network of the historical map of the VI-IV century BCE (Figure 1a) was overlapped, respectively, with the predicted clay content map and the corresponding RER map. The SW-NE narrow-shaped features with high values of clay content (low sand content), associated with the positive anomalies in Figure 8 and coincident with the riverbed of the secondary rivers Proto-Idice and Proto-Valreno, do not emerge in the RER soil texture map, but are evident in the predicted map (Figure 9b).

It is worth noticing that the hydrographic setting of the historical map used for the comparison is supported by the geo-morphological features map provided by the RER Geological Service, which also highlights that the areas with low clay content (and high sand content) correspond to the traces of sand dunes (Figure 9).

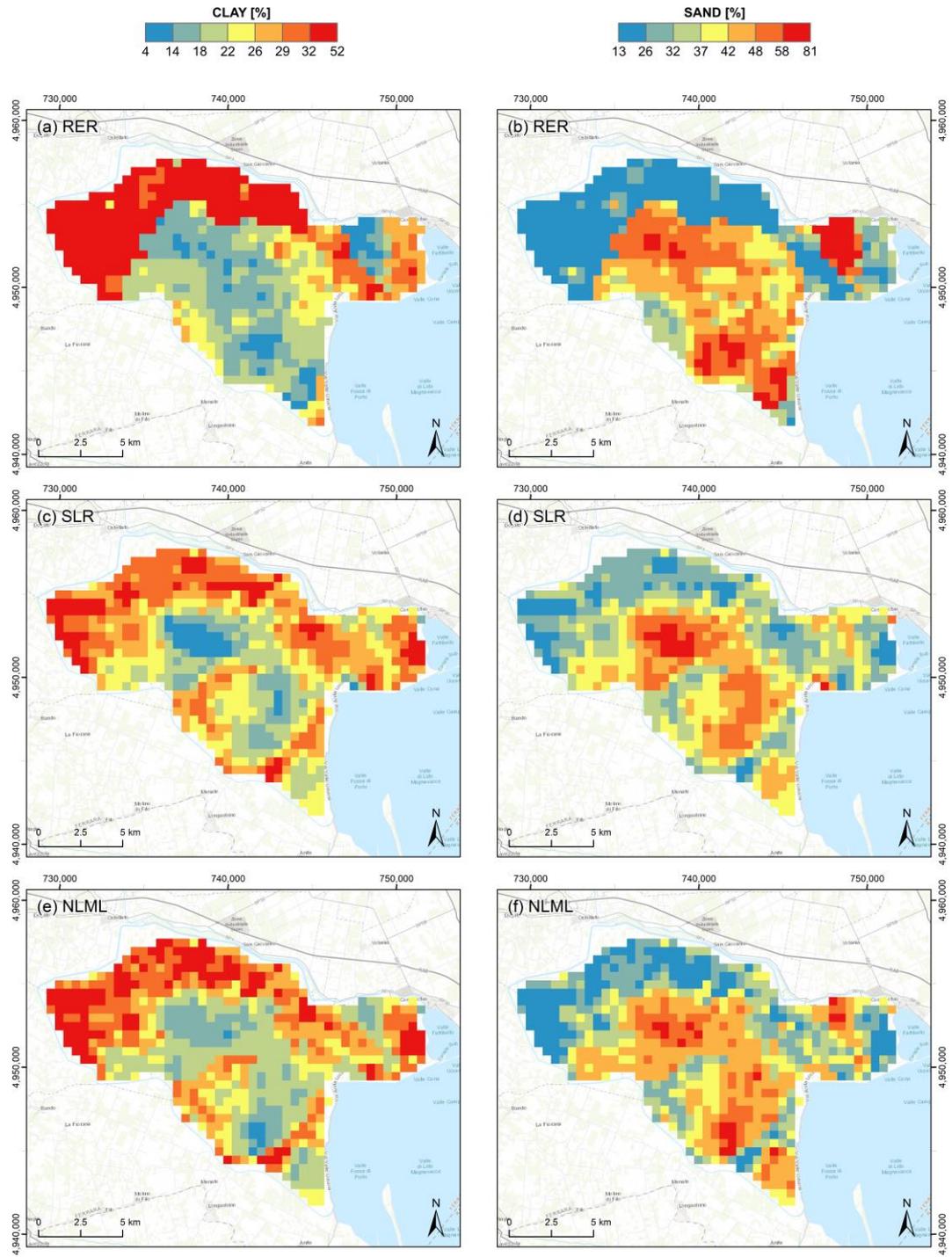

**Figure 7.** Clay maps provided by RER (**a**), predicted SLR models starting from K abundances (**c**), and by the NLML algorithm (**e**). Sand maps provided by RER (**b**), predicted by SLR models starting from K abundances (**d**) and by the NLML algorithm (**f**). Cartographic reference system: WGS 84, UTM Zone 32N.

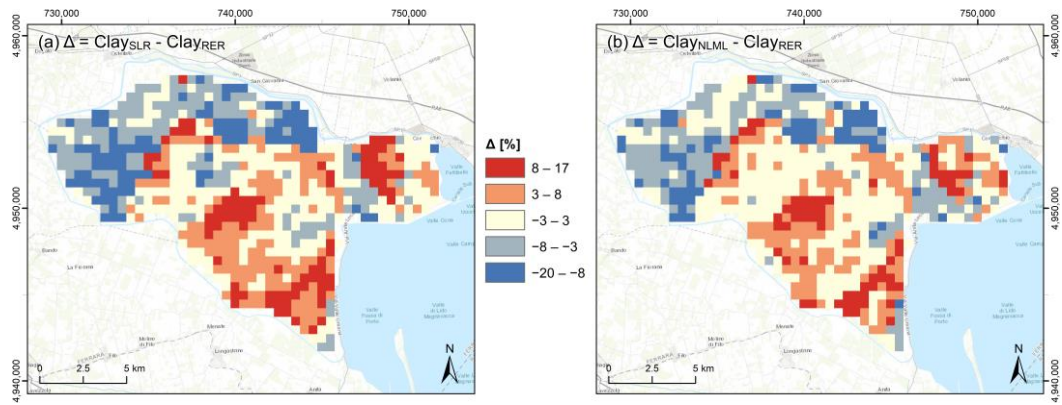

**Figure 8.** (**a**) Map of the difference between clay values predicted by the SLR model starting from K abundances and clay values reported in the RER soil textural map (Clay$_{SLR}$–Clay$_{RER}$); (**b**) map of the differences between clay values predicted by the NLML model and clay values reported in the RER soil textural map (Clay$_{NLML}$–Clay$_{RER}$). Cartographic reference system: WGS 84, UTM Zone 32N.

The correspondence between high predicted clay content areas and paleo-channel traces could be easily explained by the abandonment process of their former riverbeds, in this case, attributable to avulsion phenomena (upstream shifting of a channel in correspondence with bifurcation). Such processes are associated with fining upward sedimentation, unlike artificial fluvial diversions that leave in the abandoned riverbeds coarse sediments compatible with the high energy of the fluvial currents. In the case of avulsion, abandoning channels experience an extended transitional stage with shallowing and narrowing, with the grain size of deposits gradually fining upwards from sand to silt, and a final stage of full disconnection (Toonen, Kleinhans and Cohen [11]). This eventually completes the sequence with sediments representing the suspended load brought during floods, consisting of much finer-grained deposits (typically clay). In the Mezzano Lowland, the preservation of these sedimentary sequences was favored by the spread of salt marshes, which left the area undisturbed after the abandonment of the Eridano, Idice and Valreno.

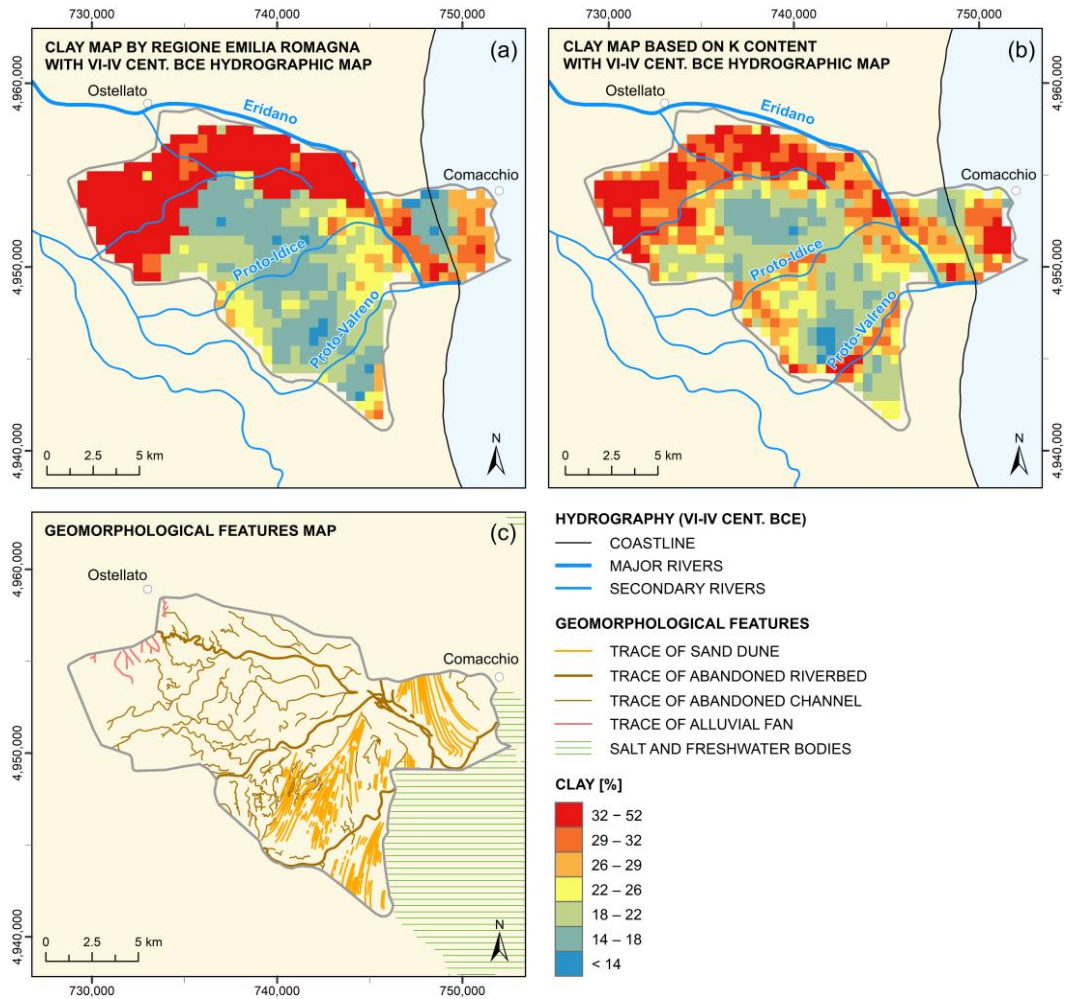

**Figure 9.** Hydrographic network of the VI-IV century BCE [30] superimposed on (**a**) the clay content map provided by RER and (**b**) the clay content map as predicted by the NLML model. (**c**) The geo-morphological feature map provided by RER. Cartographic reference system: WGS 84, UTM Zone 32N.

## 4. Conclusions

In this work, we investigated the correlations between K and Th abundances obtained from the geo-localized spectra acquired via AGRS and soil texture data extracted from the RER soil texture map of the Mezzano Lowland area. The analysis was performed with three different approaches (SLR—simple linear regression, MLR—multiple linear regression and NLML—non-linear machine learning) using 80% of the data for model training and 20% of the data for performance testing. The obtained results in terms of soil texture maps were interpreted using historical information on the hydrographic evolution of the area. We summarize here the main outcomes of this study.

- The results of the SLR analysis highlighted moderate negative correlations between sand and K abundances (r = −0.62) and between sand and Th abundances (r = -0.56). The intercepts of both regression lines are compatible at the 2σ level with a soil sand content of 100%, corresponding to null K and Th abundances. These results corroborate the presence of sandy soils with low radioactivity and high silica content in the Mezzano Lowland.

- The high cation exchange capacity of clay minerals is confirmed by a positive correlation between clay and K (Th) abundances with r = 0.64 (r = 0.53). This trend, also supported by the MLR model, permitted the production of a map of the clay content derived from radiometric data to be compared with the RER soil texture map.

- The models based on the NLML algorithm show the best performances, in terms of $R^2$, in the prediction of clay and sand soil content from K and Th abundances.

- Because of the high density of airborne data, the investigation of the spatial distributions of the clay values differences between models' predictions and RER observations permitted uncovering detailed geo-morphological features, which are not reported in the RER soil texture map. The clay maps derived from both SLR and NLML models highlight areas with high clay content attributable to the paleo-channels known as Idice, Valreno and Eridano, which crossed the Mezzano Lowland for approximately a thousand years in the Etruscan and Roman periods.

In this study, we proved that the AGRS performances for discriminating against different texture classes could be significantly improved by implementing machine learning techniques by exploiting the availability of large amounts of soil texture data to train non-linear algorithms. It is worth highlighting that AGRS data can be affected by systematic uncertainties coming from spectral noise produced by atmospheric radon [49,50], calibration processes [51,52] and soil water content [53]. Since these systematic uncertainties are very sensitive to varying environmental conditions, an uncritical application of the correlation models must be made cautiously. The greater the accuracy of the gamma measurements, the greater the reliability of the AGRS method in retrieving soil texture information.

Soil digital mapping plays a decisive role in the sustainable management of soil, a non-renewable resource increasingly affected by the consequences of climate change. From this perspective, the promising predictions given by the proposed methodology can support paleo-hydrography studies, the rational planning of agricultural practices, and the analysis of soil degradation processes.

**Author Contributions:** Conceptualization, A.M., F.M., K.G.C.R., F.S. and V.S.; methodology, A.M., F.M., K.G.C.R., F.S. and V.S.; software, A.M., M.A., E.C., M.M. (Michele Montuschi) and F.S.; formal analysis, A.M., M.A., E.C., M.M. (Michele Montuschi), K.G.C.R. and F.S.; writing—original draft, A.M., M.A., F.S. and V.S.; writing—review and editing, A.M., E.C., T.C., E.G., N.L., F.M., M.M. (Michele Montuschi), K.G.C.R., F.S. and V.S.; visualization, L.C., M.D.C., N.L., M.M. (Maurizio Marcialis), K.G.C.R. and F.S.; supervision, F.M. and V.S.; investigation, A.M., M.A., E.C., F.M., K.G.C.R. and V.S.; resources, E.A., M.M. (Maurizio Marcialis), N.M., S.P. and A.R.; data curation, A.M., M.A., E.C., M.M. (Michele Montuschi), K.G.C.R. and F.S.; project administration, E.A., T.C., E.G., F.M. and V.S.; funding acquisition, E.A., L.C., T.C., M.D.C., E.G., F.M., N.M., S.P. and A.R.; validation, A.M., F.M. and V.S. All authors have read and agreed to the published version of the manuscript.

**Funding:** This work was partially founded by: (i) ITALian RADioactivity project (ITALRAD) of the National Institute of Nuclear Physics (INFN), (ii) "TOMato for baby food: Monitoring heavY metal in production chain" project - CUP: C66B20001120008, (iii) "Monitoraggio degli sversamenti illegali attraverso l'impiego sinergico di tecnologie avanzate–C4E" project - CUP B46C18000750005, (iv) FAR 2020–2021 of the University of Ferrara and (v) GEOexplorer Impresa Sociale S.r.l..

**Data Availability Statement:** The data that support the findings of this study are available from the corresponding author, A.M., upon reasonable request.

**Acknowledgments:** The authors thank Maurino Antongiovanni, Paolo Baldassarri, Stefano Calderoni, Adriano Facchini, Claudio Pagotto, and Andrea Serafini for their valuable support. This paper is dedicated to Prof. Giovanni Fiorentini who passed away in June 2022. He was an extraordinary mentor for students and young scientists: without his inspiration and vision this study would not have been possible. The members of the Laboratory for Nuclear Technologies Applied to the Environment are infinitely grateful to Prof. Giovanni Fiorentini.

**Conflicts of Interest:** The authors declare no conflicts of interest.